\documentclass[aps,prb,twocolumn,superscriptaddress,showpacs]{revtex4-1}

\usepackage{graphicx}
\usepackage{calc}
\usepackage{bm}
\usepackage{color}

\begin{document}

\title{Conductance oscillations and zero-bias anomaly in a single superconducting junction to a three-dimensional $Bi_2Te_3$ topological insulator}

\author{O.O.~Shvetsov}
\affiliation{Institute of Solid State Physics RAS, 142432 Chernogolovka, Russia}
\author{V.A.~Kostarev}
\affiliation{Institute of Solid State Physics RAS, 142432 Chernogolovka, Russia}
\author{A.~Kononov}
\affiliation{Institute of Solid State Physics RAS, 142432 Chernogolovka, Russia}
\author{V.A. Golyashov}
\affiliation{A.V. Rzanov Institute of Semiconductor Physics, Siberian Branch, Russian Academy of Sciences, 630090 Novosibirsk, Russia}
\affiliation{Saint-Petersburg State University, 198504 Saint-Petersburg, Russia}
\author{K.A. Kokh}
\affiliation{Saint-Petersburg State University, 198504 Saint-Petersburg, Russia}
\affiliation{Novosibirsk State University, 630090 Novosibirsk, Russia}
\affiliation{V.S. Sobolev Institute of Geology and Mineralogy, Siberian Branch, Russian Academy of Sciences, 630090 Novosibirsk, Russia}
\author{O.E. Tereshchenko}
\affiliation{A.V. Rzanov Institute of Semiconductor Physics, Siberian Branch, Russian Academy of Sciences, 630090 Novosibirsk, Russia}
\affiliation{Saint-Petersburg State University, 198504 Saint-Petersburg, Russia}
\affiliation{Novosibirsk State University, 630090 Novosibirsk, Russia}
\author{E.V.~Deviatov}
\affiliation{Institute of Solid State Physics RAS, 142432 Chernogolovka, Russia}

\date{\today}

\begin{abstract}
We experimentally investigate Andreev transport through a single junction between an s-wave indium superconductor and a thick film of a three-dimensional $Bi_2Te_3$ topological insulator. We study $Bi_2Te_3$ samples with different bulk and surface characteristics, where the presence of a topological surface state is confirmed by direct ARPES measurements. All the junctions demonstrate Andreev transport within the superconducting gap. For junctions with transparent $In-Bi_2Te_3$ interfaces we find  a number of nearly periodic conductance oscillations, which are accompanied by zero-bias conductance anomaly. Both effects disappear above the superconducting transition or for resistive junctions. We propose a consistent interpretation of both effects as originating from proximity-induced  superconducting correlations within the $Bi_2Te_3$ topological surface state.
\end{abstract}

\pacs{74.45.+c  73.23.-b 73.40.-c}

\maketitle

\section{Introduction}

Recent interest to Andreev reflection is mostly connected with new materials, which can be characterized by Dirac spectrum. They are graphene~\cite{geim},  three-~\cite{hasan} and two-dimensional~\cite{konig,kvon} topological insulators, topological Dirac semimetals~\cite{semid,semid1}. Classical Andreev reflection~\cite{andreev} allows low-energy electron transport from normal metal to superconductor  by creating  a Cooper pair, so a hole is reflected back to the normal side of the junction~\cite{tinkham}.  The Andreev process is more complicated for relativistic materials. For example, the scattering potential~\cite{BTK} at the normal-superconductor (NS) interface do not suppress the current~\cite{yung} due to the Klein paradox~\cite{klein,klein1}. Also, the interband reflection is possible, where a reflected hole appears in the valence band, which is known as specular Andreev reflection~\cite{been1,been2}. This process has been recently reported for graphene~\cite{efetov} and for two-dimensional semimetals~\cite{nbgasb}.

Proximity-induced superconductivity is regarded to be responsible for conductance oscillations in transport along the topological surface state. Bogoliubov quasiparticles are predicted~\cite{adroguer} to  experience Fabry-Perot-type transmission resonances for energies above the superconducting gap within the  region of induced superconductivity for a long $L>>\xi$ single NS junction, $\xi$ is a coherence length.  On the other hand, subgap conductance oscillations were demonstrated~\cite{finck} for a short $L<<\xi$ three-dimensional $Bi_2Se_3$ topological insulator sandwiched between  superconducting and normal leads. These effects are the representations of Tomasch~\cite{tomasch1,tomasch2} and MacMillan-Rowell~\cite{mcmillan1,mcmillan2} geometrical resonances, which originates in classical~\cite{tomasch1,mcmillan1,mcmillan2,tomasch_exp1,tomasch_exp2} NS junctions due to the space restriction in the S or N regions, respectively.

Three-dimensional topological insulators like  $Bi_2Te_3$ are usually characterized~\cite{hasan,volkov} by both topological surface and trivial bulk transport contributions. Following Ref.~\cite{adroguer}, it seems to be reasonable to search for resonances also in charge transport through  wide $L>>\xi$ superconducting junctions to these materials, which could also help to distinguish between two transport channels. 

Here, we experimentally investigate Andreev transport through a single junction between an s-wave indium superconductor and a thick film of a three-dimensional $Bi_2Te_3$ topological insulator. We study $Bi_2Te_3$ samples with different bulk and surface characteristics, where the presence of a topological surface state is confirmed by direct ARPES measurements. All the junctions demonstrate Andreev transport within the superconducting gap. For junctions with transparent $In-Bi_2Te_3$ interfaces we find  a number of nearly periodic conductance oscillations, which are accompanied by zero-bias conductance anomaly. Both effects disappear above the superconducting transition or for resistive junctions. We propose a consistent interpretation of both effects as originating from proximity-induced  superconducting correlations within the $Bi_2Te_3$ topological surface state.

\section{Samples and technique}

\begin{figure}
\includegraphics[width=0.8\columnwidth]{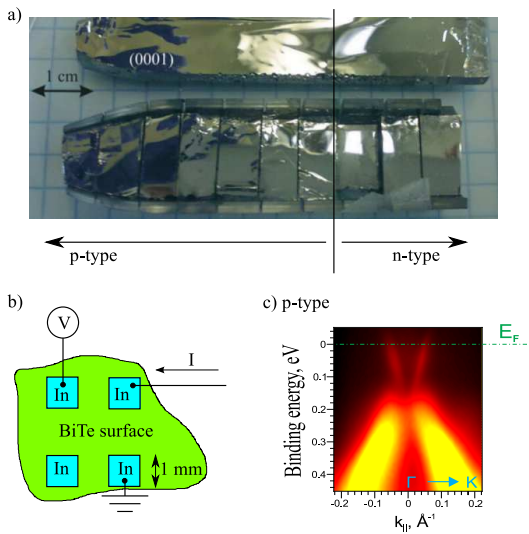}
\caption{(Color online) (a) Image of a single $Bi_2Te_3$ crystal  cleaved along (0001) surface. Because of Te segregation~\cite{bathon}, small parts of the crystal are characterized by different type of bulk carriers along the growth direction.
(b) Sketch of the sample with several indium contacts to the cleaved $Bi_2Te_3$ surface.  Charge transport is investigated in a three-point technique: one superconducting In electrode is grounded; two others are used to feed a current $I$ and to measure the potential $V$. (c)  ARPES data for $Bi_2Te_3$ (0001) surface along  the $\Gamma$K direction ($h\nu$= 23~eV, T=300~K).  The topological surface state and the topmost bulk valence band can be seen for the p-type crystal.
}
\label{sample}
\end{figure}

Single crystals of $Bi_2Te_3$ were grown using the modified Bridgman method, see Ref.~\cite{teresch} for details. It was shown~\cite{bathon}, that solidification of 61 or 62 mol.\% Te melts provides transition from regions of p- to n- type bulk carriers along the growth direction, which is caused by Te segregation.  Apart from the narrow compensated region, both p- and n-parts of the crystal show high low-temperature (4 K) carrier mobility $\approx 10^{4}  $cm$^{2}$/Vs for carrier concentration $\approx  10^{19}  $cm$^{-3}$.

Different samples are exfoliated from  p- or n-type parts of a single $Bi_2Te_3$ crystal, see in Fig.~\ref{sample} (a). Exfoliation produces thick (below 1~$\mu$m) $Bi_2Te_3$ films with freshly cleaved (0001) surface. Superconducting contacts are made by thermal evaporation of 100~nm  indium  through the shadow mask on this surface, see the sample sketch in Fig.~\ref{sample} (b). 

The evaporated In film is checked to undergo superconducting transition at $T_c=3.4~K$, in good accordance with the known literature data~\cite{kittel}. At lower temperatures, every $Bi_2Te_3 - In$ contact represents single NS junction of macroscopic, $\approx 1$~mm, lithographic dimensions, which exceeds well the indium correlation length~\cite{indium} $L>>\xi_{In}\approx 300$~nm. The latter requirement is obligatory  within the framework of Ref.~\cite{adroguer}, so we use thick $Bi_2Te_3$ films of macroscopic dimensions, instead of the commonly used micrometer-size thin flakes~\cite{finck}.

We study charge transport across one particular NS junction in a three-point technique, as depicted in Fig.~\ref{sample} (b): one superconducting In contact is grounded; two others are used to feed a current $I$ through the junction and to measure the potential $V$. 
In general, $V$  reflects the in-series connected resistances of the grounded $Bi_2Te_3 - In$ junction  and some part of $Bi_2Te_3$ crystal between the ground and potential contacts, see Fig.~\ref{sample} (b). If the former term is dominant, the obtained $I-V$ curve should be independent of  the particular  choice of current/voltage probes, which is the experimental test on relative bulk influence. 

We sweep the dc current  from -50~$\mu$A to +50~$\mu$A. To obtain $dV/dI(V)$ characteristics,  the dc current is modulated by a low (0.1~$\mu$A) ac (110~Hz) component. We measure both the dc ($V$) and ac ($\sim dV/dI$) components of the $Bi_2Te_3$ surface potential by using a dc voltmeter and a lock-in amplifier, respectively. The lock-in signal is independent of the modulation frequency up to 1~kHz, which is defined by applied ac filters. To investigate Andreev transport, the measurements are performed below $T_c$.

One can expect a complicated transport behavior for the $Bi_2Te_3-In$ NS junction, since  topological surface state is present at the $Bi_2Te_3$ surface. Fig.~\ref{sample} (c) shows typical ARPES data obtained at T=300~K on the vacuum cleaved (0001) surface of p-type $Bi_2Te_3$ crystals  along the $\Gamma$K direction at a photon energy of $h\nu$= 23~eV. The topological surface state and the topmost bulk valence band can be seen in Fig.~\ref{sample} (c).

\section{Experimental results}

\begin{figure}
\includegraphics[width=\columnwidth]{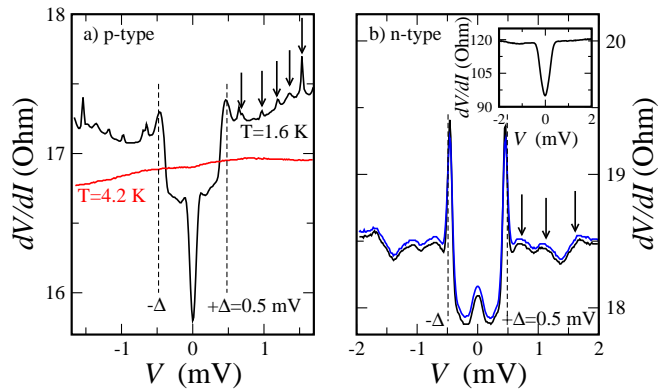}
\caption{(Color  online) Examples of  $dV/dI(V)$ characteristics for $Bi_2Te_3-In$ NS junctions with p- (a) and n-type (b) bulk $Bi_2Te_3$ conductivity.  At low $T=1.6$~K, Andreev reflection can be seen as a wide drop in $dV/dI$ for biases $|eV|<\Delta_{In}=0.5$~mV, which is  suppressed at high temperature $T=4.2$~K (the red curve  in the (a) part). The low-energy transport behavior  appears as a zero-bias $dV/dI$ peak or dip. For both junction types, we observe nearly periodic  $dV/dI$ oscillations for $|eV|>\Delta_{In}$, which are perfectly symmetric in respect to the bias sign. The positions of the oscillations are denoted by arrows.  Two  curves in the (b) part differ by the current/voltage probes positions, the effect on $dV/dI(V)$ is negligible. Inset demonstrates an example of $dV/dI(V)$ curve for the resistive NS junction, which is characterized by high single-particle scattering at the NS interface. Both  the oscillations and the zero-bias anomaly can not be seen.
} 
\label{IV}
\end{figure}

Fig.~\ref{IV} presents typical examples of $dV/dI(V)$ characteristics. They demonstrate behavior, which is qualitatively consistent with  standard Andreev one for single SN junction~\cite{tinkham}. At high temperature $T=4.2$~K, $dV/dI$ differential resistance is nearly linear, see Fig.~\ref{IV} (a). At low $T=1.6$~K, there is a wide drop in $dV/dI$ within $eV\approx \pm 0.5$~mV bias region, see Fig.~\ref{IV} (a) and (b), which corresponds well to the indium superconducting gap~\cite{kittel} $\Delta_{In}=0.5$~mV. According to standard BTK theory~\cite{BTK}, the drop in $dV/dI$ within the superconducting gap is a fingerprint of Andreev reflection at the transparent  NS interface~\cite{tinkham}.

In general, $dV/dI(V)$ curves could contain some admixture of bulk $Bi_2Te_3$  resistance, as it can be seen from Fig.~\ref{sample} (b). However, at given concentrations and mobilities, a characteristic bulk resistance value is much below 1 Ohm. This estimation is experimentally verified  by changing the current/voltage probes in Fig.~\ref{IV} (b).  Thus, the $dV/dI(V)$ curves in Fig.~\ref{IV} reflect only the $In-Bi_2Te_3$ junction transport characteristics. 

Surprisingly, we observe pronounced nearly periodic $dV/dI$ oscillations for biases above the superconducting gap $|eV|>\Delta_{In}$, see Fig.~\ref{IV} (a) and (b). The oscillations are  symmetric in respect to the bias sign and are seen in a wide bias range.

The $dV/dI(V)$ curves also demonstrate the zero-bias transport anomaly, which appears as a $dV/dI$ peak  or  dip, see Fig.~\ref{IV} (a) and (b). We wish to mention here, that both the oscillations and the zero-bias anomaly appear together for our samples: they can not be seen for resistive NS junctions, see the inset to Fig.~\ref{IV} (b) or at high temperature $T>T_c$, so they require consistent  explanation.

\begin{figure}
\centerline{\includegraphics[width=0.95\columnwidth]{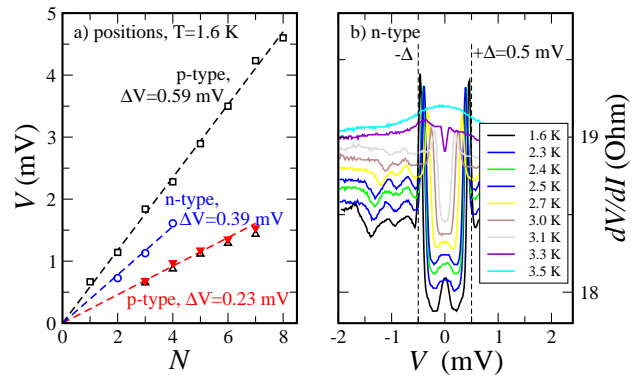}}
\caption{(Color  online) (a) The positions of the oscillations vs. their numbers for: the same $Bi_2Te_3-In$ junction in two different cooling cycles (up and down triangles, the dashed line reflects $\Delta V=0.23$~mV period), $dV/dI(V)$ for this junction is presented in Fig.~\protect\ref{IV} (a); another p-type junction (squares, $\Delta V=0.59$~mV); n-type junction (circles, $\Delta V=0.39$~mV), $dV/dI(V)$ is in Fig.~\protect\ref{IV} (b). (b) Temperature dependence of $dV/dI(V)$ for the n-type $Bi_2Te_3-In$ junction. The wide $|eV|<\pm$0.5~mV  $dV/dI$  drop and the oscillations disappear together exactly above  the indium $T_c=3.4$~K. The zero-bias $dV/dI$ peak disappears earlier, between 2.7~K and 3~K.
} 
\label{IV_T}
\end{figure}

The positions of the oscillations are depicted in Fig.~\ref{IV_T} (a). Experimental points cover wide bias range  for biases much above $\Delta_{In}$. The data demonstrate clear linear dependencies, so the oscillations are  periodic in respect to the bias.  As it can be seen from Fig.~\ref{IV_T} (a), the period $\Delta V$ is not universal, it is different for different samples. On the other hand, for a particular sample $\Delta V$ is well reproduced in different cooling cycles.

Fig.~\ref{IV_T} (b) demonstrates temperature dependence of $dV/dI(V)$ characteristics.  The  wide  $|eV|<\pm$0.5~mV  drop in $dV/dI$ disappears exactly above the indium $T_c=3.4$~K, as it is expected for Andreev reflection~\cite{tinkham}.  The amplitude of the oscillations is  suppressed by  temperature together with the superconducting gap. On the other hand, $dV/dI$ change is small near $T_c^{In}$ in Fig.~\ref{IV_T} (b). Thus, despite we observe  the transport oscillations for $|eV|>\Delta_{In}$, they are also induced by superconductivity in the  $Bi_2Te_3-In$ junction.  The zero-bias $dV/dI$ peak disappears earlier, between 2.7~K and 3.0~K.

\section{Discussion}

\begin{figure}
\includegraphics[width=0.8\columnwidth]{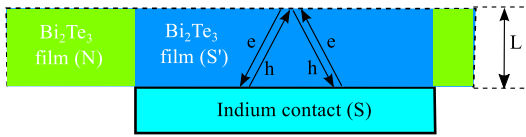}
\caption{(Color online)  Schematic diagram of scattering between the $Bi_2Te_3-In$  interface and  the opposite surface of the $Bi_2Te_3$ film. The  Andreev-reflected particle experience normal scattering back to the $Bi_2Te_3-In$ interface, as depicted by arrows. Because of the particle type change in Andreev reflection, two Andreev reflections  are required  for the interference to occur in the film. The scattering is specular on the opposite (normal) film surface, but we suppose usual retro-reflection at the NS interface. Electron and hole trajectories are shifted for clarity. The region of induced superconducting correlations S' is schematically shown by blue color to be different from normal $Bi_2Te_3$ regions (yellow).
}
\label{discussion}
\end{figure}

The oscillating behavior requires scattering events at two different interfaces. For example, multiple Andreev reflections~\cite{tinkham,MAR}  are observed in the SNS structures with two superconducting leads. Periodic resonances in a single NS junction appear if there is a space restriction in the superconducting (Tomasch effect~\cite{tomasch1,tomasch2}) or normal (MacMillan-Rowell one~\cite{mcmillan1,mcmillan2}) side of the junction. In the former case~\cite{tomasch1,tomasch2}, resonances result from Bogoliubov quasiparticle interferences between an incident electron-like quasiparticle and its hole-like counterpart. The nearly periodic resonances appear for energies above the superconducting gap with  $\approx \pi\hbar v_F/2L$ period, where $v_F$ is the Fermi velocity in the S and $L$ its thickness. In the latter case~\cite{mcmillan1,mcmillan2}, the Andreev-reflected particle experience normal scattering back to the NS interface. Because of the particle type change in Andreev reflection, two reflections  are required at the NS interface for the interference to occur in the N, see Fig.~\ref{discussion}. Thus, the resonances appears with $\pi\hbar v_F/4L$ period, if superconducting correlations survive over $L$. 

Tomasch oscillations within the indium film can not be responsible for our experimental results: monotonous $dVdI(V)$ curves are obtained for resistive NS junctions, see an example in the inset to Fig.~\ref{IV} (b). Because of the same indium film thickness and quality, the resonances are not connected with the space restriction in the S side of the $Bi_2Te_3-In$  junction. 

On the other hand, there are no oscillations above $T_c$ in Fig.\ref{IV_T} (b), so the oscillations should be connected with induced superconductivity  within  $Bi_2Te_3$ films.

(i) A variant of Ref.~\cite{adroguer} for oscillations within the region of proximity-induced superconductivity within the topological surface state is qualitatively appropriate. The Fermi velocity can be directly obtained for the topological surface state from the ARPES data in Fig.~\ref{sample} (c) as $v_F\approx 4.8\times10^{7}$ cm/s. If we take the minimal value of experimental period in Fig.~\ref{IV_T} (a)  $\Delta V=0.23$~mV, the maximum $2L$ value can be estimated as $2L \sim \pi\hbar v_F/ e\Delta V \approx 4 \mu$m. The discrepancy with the lithographic value (1~mm) requires strongly inhomogeneous transport, which goes through some region of width $L$ in wide $Bi_2Te_3-In$  contact area. This is supported by $dV/dI$ values  in Fig.~\ref{IV}: every $dV/dI(V)$ curve is still a wide $dV/dI$ drop within the indium superconducting gap, even  in  the inset to Fig.~\ref{IV}, which implies highly transparent NS interfaces in standard BTK model~\cite{BTK}. In this case, the resistance of 1~mm size junction should be well below 1~Ohm, which obviously contradicts to that we have in Fig.~\ref{IV}.

(ii) Another possibility is the MacMillan-Rowell oscillations for bulk carriers in  $Bi_2Te_3$ films, which appears due to the finite film thicknesses. In this case, a bulk particle is scattered between the NS interface and the opposite surface of the film as schematically depicted in Fig.~\ref{discussion}. The $4L$ distance can be estimated as   $\approx 4 \mu$m for the minimal $\Delta V=0.23$~mV, while it is approximately twice smaller ($1.6 \mu$m) for the maximal $\Delta V=0.59$~mV. The obtained $L$ values for different samples  are in good correspondence with the estimations made in exfoliation process. Also, they are much below the mean free path, which can be estimated as ($l_e\approx 60 \mu$m for given concentrations and mobility~\cite{teresch,bathon}. 

As a result, interpretation of periodic $dV/dI$ oscillations for finite biases may support topological effects, but does not obligatory require them. The situation is different for the zero-bias anomaly in transparent NS junctions. 

The zero-bias anomaly is known for Andreev transport in different regimes. It was previously connected with disorder effects~\cite{diar} or with more exotic scenarios like Majorana fermions~\cite{Heiblum,Mourik,Deng} and Andreev surface states~\cite{ass}. It also appears for ballistic transport $l_e>> L$, if there is some restricted area within N region~\cite{heslinga,klapwijk17,akhmerov17}. In our samples, the topological surface state is present at the $Bi_2Te_3$ surface, see Fig.~\ref{sample} (c), being separated from the bulk by depletion region.   The depletion is usually formed~\cite{zav1,zav2,hofman} near the surface due to carriers' localization  by adsorbates, heterointerface formation~\cite{depletion}, and defects, it can vary for different junctions. Thus, it is natural to connect the zero-bias anomaly with the superconductivity, induced in the topological surface state on the $Bi_2Te_3$ surface. Moreover, it  explains why the oscillations and the zero-bias anomaly appear together in our samples, as it can be seen in  Fig.~\ref{IV}.

As for numerical estimations, the gap $\Delta_{ind}$ induced within the surface state can be obtained from the width of the low-bias structure~\cite{heslinga,klapwijk17} as $0.1$~meV in Fig.~\ref{IV} (a) and as $0.17$~meV in Fig.~\ref{IV} (b). It should be defined by Thouless energy $\Delta_{ind} \sim E_{Th}$ (see Appendix to Ref.~\cite{akhmerov17} for recent comprehensive discussion), which is defined as  $ E_{Th}\sim\hbar v_F/L$ in the regime of ballistic transport~\cite{golubov15}. Obviously, it gives $L$ values of the same order that we have from the oscillation periods in Fig.\ref{IV_T} (a). Which is more important, the width of the zero-bias anomaly scales for different junctions exactly as the period of the oscillations (0.17/0.1=0.39/0.23 in Fig.~\ref{IV} (a) and (b)), which strongly supports our interpretation of the zero-bias anomaly. Also, as expected for $E_{Th}$,  it is constant in Fig.~\ref{IV_T} (b), until $k_BT$  exceeds $\Delta_{ind}$.

\section{Conclusion}

As a conclusion, we experimentally investigate Andreev transport through a single junction between an s-wave indium superconductor and a thick film of a three-dimensional $Bi_2Te_3$ topological insulator. We study $Bi_2Te_3$ samples with different bulk and surface characteristics, where the presence of a topological surface state is confirmed by direct ARPES measurements. All the junctions demonstrate Andreev transport within the superconducting gap. For junctions with transparent $In-Bi_2Te_3$ interfaces we find  a number of nearly periodic conductance oscillations, which are accompanied by zero-bias conductance anomaly. Both effects disappear above the superconducting transition or for resistive junctions. We propose a consistent interpretation of both effects as originating from proximity-induced  superconducting correlations within the $Bi_2Te_3$ topological surface state.

\acknowledgments
We wish to thank Ya.~Fominov, D.E.~Feldman, V.T.~Dolgopolov, and T.M.~Klapwijk for fruitful discussions. We gratefully acknowledge financial support by the RFBR (project No.~16-02-00405), RAS, and Saint Petersburg State University (Grant Number 15.61.202.2015). O.E.T. acknowledge financial support by the Russian Science Foundation (project n. 17-12-01047), in part of crystal growth, structural characterization, and ARPES measurements (Fig.1 (c)).


\begin{thebibliography}{99}


\bibitem{geim} K.S. Novoselov, A.K. Geim, S. Morozov, D. Jiang, Y. Zhang, S. Dubonos, I. Grigorieva, A. Firsov, Science 306, 666 (2004).
\bibitem{hasan} M. Z. Hasan and C. L. Kane, Rev. Mod. Phys. 82, 3045 (2010); D. Hsieh, D. Qian, L. Wray, Y. Xia, Y. S. Hor, R. J. Cava,
and M. Z. Hasan, Nature (London) 452, 970 (2008).
\bibitem{konig} M. K\"onig, S. Wiedmann, C. Br\"une, A. Roth, H. Buhmann, L. W. Molenkamp, X.-L. Qi, and S.-C. Zhang, Science 318, 766 (2007).
\bibitem{kvon} G. M. Gusev, Z. D. Kvon, O. A. Shegai et al. , Phys. Rev. B 84, 121302(R) (2011).
\bibitem{semid} Z. Wang, Y. Sun, X.-Q. Chen, C. Franchini, G. Xu, H. Weng, X. Dai, Z. Fang, Phys. Rev. B 85, 195320 (2012).
\bibitem{semid1} S.M. Young, S. Zaheer, J.C. Teo, C.L. Kane, E.J. Mele, A.M. Rappe,  Phys. Rev. Lett. 108, 140405 (2012).

\bibitem{andreev} A. F. Andreev, Soviet Physics JETP 19, 1228 (1964).
\bibitem{tinkham} M. Tinkham, Introduction to Superconductivity (2d ed., McGraw–Hill, New York, 1996).

\bibitem{BTK} G.E. Blonder, M. Tinkham, T.M. Klapwijk, Physical Review B, 25, 4515 (1982).
\bibitem{yung} arxiv:1701.00591
\bibitem{klein} M. I. Katsnelson, K. S. Novoselov, and A. K. Geim, Nat. Phys. 2, 620 - 625 (2006).
\bibitem{klein1}  C. W. J. Beenakker, Rev. Mod. Phys. 80, 1337–1354 (2008).

\bibitem{been1} C. W. J. Beenakker, Physical Review Letters 97 (2006).
\bibitem{been2} C. W. J. Beenakker, Reviews of Modern Physics 80, 1337 (2008).
\bibitem{efetov} D. K. Efetov, L. Wang, C. Handschin, K. B. Efetov, J. Shuang, R. Cava, T. Taniguchi, K. Watanabe, J. Hone, C. R. Dean, P. Kim,  Nature Physics (2015);     doi:10.1038/nphys3583
\bibitem{nbgasb} A. Kononov, S. V. Egorov, V. A. Kostarev, B. R. Semyagin, V. V. Preobrazhenskii, M. A. Putyato, E. A. Emelyanov, and E. V. Deviatov, JETP Letters,  Vol. 104, No. 1, pp. 26–31 (2016).

\bibitem{adroguer} P. Adroguer, C. Grenier, D. Carpentier, J. Cayssol, P. Degiovanni, and E. Orignac, Phys. Rev. B 82, 081303(R), (2010).
\bibitem{finck} A.D.K. Finck, C. Kurter, Y.S. Hor, D.J. Van Harlingen,  Phys. Rev. X 4, 041022 (2014)

\bibitem{tomasch1} W. J. Tomasch, Phys. Rev. Lett. 16, 16 (1966).
\bibitem{tomasch2}  W. L. McMillan, P. W. Anderson, Phys. Rev. Lett. 16, 85 (1966).
\bibitem{mcmillan1} J. M. Rowell, W. L. McMillan, Phys. Rev. Lett. 16, 453 (1966).
\bibitem{mcmillan2} J. M. Rowell,  Phys. Rev. Lett. 30, 167–170 (1973).
\bibitem{tomasch_exp1}  O. Nesher, G. Koren, Phys. Rev. B 60, 9287 (1999).
\bibitem{tomasch_exp2} C. Visani, Z. Sefrioui, J. Tornos, C. Leon, J. Briatico, M. Bibes, A. Barth\'el\'emy,
J. Santamaría and Javier E. Villegas,  Nature Physics, 8, 539 (2012), DOI: 10.1038/NPHYS2318

\bibitem{volkov} B.A. Volkov, O.A. Pankratov, JETP Letters 42, 178 (1985).



\bibitem{teresch} K.A. Kokh, S.V. Makarenko, V.A. Golyashov, O.A. Shegai, O.E. Tereshchenko, CrystEngComm., 16, 581 (2014).
\bibitem{bathon} T. Bathon, S. Achilli, P. Sessi, V.A. Golyashov, K. A. Kokh, O. E. Tereshchenko, and M. Bode,  Advanced Materials 28, issue 11, 2183–2188 (2016). doi:10.1002/adma.201504771

\bibitem{kittel} Charles Kittel. Introduction to Solid State Physics, 5th edition. New York: John Wiley \& Sons, Inc, 1976.
\bibitem{indium} A. M. Toxen, Phys. Rev. 123, 442 (1961).







\bibitem{MAR} As a recent example, see H. Y. G\"unel, N. Borgwardt, I. E. Batov, et. al.,  Nano Lett., 14 (9), 4977 (2014).

\bibitem{diar} D. I. Pikulin, J. P. Dahlhaus, M. Wimmer, H. Schomerus and C. W. J. Beenakker, New J. Phys. 14, 125011 (2012).
\bibitem{Heiblum} A. Das, Y. Ronen, Y. Most, Y. Oreg, M. Heiblum, and H. Shtrikman,  Nature Physics 8, 887 (2012)   
\bibitem{Mourik} V. Mourik, K. Zuo, S. M. Frolov, S. R. Plissard, E. P. A. M. Bakkers, L. P. Kouwenhoven, Science 336, 1003 (2012).
\bibitem{Deng} M. T. Deng, C. L. Yu, G. Y. Huang, M. Larsson, P. Caroff, H. Q. Xu  Nano Lett. 12, 6414 (2012).
\bibitem{ass} A.P. Schnyder and S. Ryu,  Phys. Rev. B 84, 060504(R) (2011).

\bibitem{heslinga} D.R.~Heslinga, S.E.~Shafranjuk, H.~van~Kempen, and T.M.~Klapwijk, Phys. Rev. B {\bf 49}, 10484 (1994).
\bibitem{klapwijk17}  J. Wiedenmann, E. Liebhaber, J.s K\"ubert, E. Bocquillon, Ch. Ames, H. Buhmann, T.M. Klapwijk, L.W. Molenkamp,  	arXiv:1706.01638.
\bibitem{akhmerov17} T.{\"O}.~Rosdahl, A.~Vuik, M.~Kjaergaard, A.R.~Akhmerov, arXiv:1706.08888v1.


\bibitem{zav1} Vul B. M. , Zavaritskaya E. I., Zavaritskii V. N., JETP Letters, 27, 547 (1978) 
\bibitem{zav2} Vul B.M., Zavaritskaya E.I.,  JETP Letters, 35, 259 (1982), 
\bibitem{hofman} Marco Bianchi, Richard C. Hatch, Jianli Mi, Bo Brummerstedt Iversen, and Philip Hofmann Phys. Rev. Lett. 107, 086802 (2011)

\bibitem{depletion} From ARPES for our samples, see an example in Fig.~\ref{sample} (c), the top of the valence band is located at a binding energy of about 170~meV for p-type samples. Taking into account the bulk Fermi level position  at the top of the valence band for p-type samples~\cite{bathon,atuchin}, this indicates strong, 170~meV, downward band bending at the $Bi_2Te_3$ interface with vacuum.

\bibitem{atuchin}  V.V. Atuchin, V.A. Golyashov, K.A. Kokh, I.V. Korolkov, A.S. Kozhukhov, V.N. Kruchinin, I.D. Loshkarev, L.D. Pokrovsky, I.P. Prosvirin, K.N. Romanyuk, O.E. Tereshchenko,  Journal of Solid State Chemistry 236, 203 (2016).
\bibitem{golubov15} M.~Snelder, M.P.~Stehno, A.A.~Golubov, C.G.~Molenaar, T.~Scholten, D.~Wu, Y.K.~Huang, W.G.~van~der~Wiel, M.S.~Golden, and A.~Brinkman, arXiv:1506.05923.

\end{thebibliography}
\end{document}